\documentclass[prl,aps,twocolumn,floatfix,showpacs]{revtex4}
\usepackage{graphicx,graphics,epsfig,epstopdf,amsmath,calc}
\begin{document}
\title{Superfluid Phases of Dipolar Fermions in Harmonically Trapped Optical Lattices}
\author{Doga Murat Kurkcuoglu, Li Han, and C. A. R. S{\'a} de Melo}
\affiliation{School of Physics, Georgia Institute of Technology, 
Atlanta, Georgia 30332, USA}
\date{\today}

\begin{abstract}
We describe the emergence of superfluid phases of ultracold dipolar 
fermions in optical lattices for two-dimensional 
systems.  Considering the many-body screening
of dipolar interactions at intermediate and larger filling factors, 
we show that several superfluid phases with distinct pairing symmetries 
naturally arise in the singlet channel: local $s$-wave $(sl)$, 
extended $s$-wave $(se)$, $d$-wave $(d)$
or time-reversal-symmetry breaking $(sl + se \pm id)$-wave.
We obtain the temperature versus filling factor phase diagram and show
that $d$-wave pairing is favored near half-filling, that $(sl + se)$-wave is
favored near zero or full filling, and that time-reversal-breaking
$(sl + se \pm id)$-wave is favored in between. The inclusion of a 
harmonic trap reveals that a sequence of phases can coexist in the cloud 
depending on the filling factor at the center of the trap.
Most notably in the spatial region where the $(sl + se \pm id)$-wave 
superfluid occurs, spontaneous currents are generated, and may be 
detected using velocity sensitive Bragg spectroscopy.
\end{abstract}
\maketitle

%
%
%
%
%
%

%
%

Ultracold heteronuclear molecules are very interesting quantum
systems to study because they possess electric dipole moments. This
internal degree of freedom adds richness to the nature of
interactions between molecules in comparison to interactions between
atoms in purely atomic systems. Dipolar molecules can be either
fermionic or bosonic in nature depending on their 
constituent atoms, and dipolar interactions allow for the
emergence of quantum phases which may be extremely difficult to be
realized in condensed matter. Recently, ultracold dipolar molecules
were produced optically~\cite{demille-2005} followed by their
production from Bose-Fermi mixtures of ultracold atoms first in the
vicinity of Feshbach resonances~\cite{ospelkaus-2006}, and later
brought into their rovibrational ground
state~\cite{ospelkaus-2010}. The production of these heteronuclear
molecules in harmonic traps has paved the way for studies of the
quantum phases of interacting dipolar bosonic of fermionic
molecules. In the case of trapped clouds, a few quantum phases have
been proposed for dipolar bosons including ferroelectric
superfluids~\cite{iskin-2007}, Wigner crystals~\cite{zoller-2007},
while for dipolar fermions phases such as 
ferroeletric~\cite{iskin-2007} or ferro-nematic~\cite{fradkin-2009}
Fermi liquids and Berezinskii-Kosterlitz-Thouless~\cite{taylor-2008}
or $j$-triplet~\cite{wu-2010} superfluids have been suggested. In
the case of optical lattices, additional phases such as
supersolids~\cite{danshita-2009} or
micro-emulsions~\cite{pollet-2009} have been proposed for dipolar
bosons, while studies for dipolar fermions in optical lattices are
just beginning.

A natural next step for experiments is the loading of dipolar
(heteronuclear) molecules in optical lattices. We have particularly
in mind fermionic dipolar molecules, such as $^{23}{\rm Na} ^{40}{\rm K}$, 
which seem to be stable against chemical reactions in their 
electronic-roto-vibrational ground-state~\cite{hutson-2010}. 
Such molecules also have hyperfine structure due to the nuclear spins, 
and a mixture of two hyperfine states with sufficient long lifetimes 
may be created~\cite{NaK-molecules-2013}.
In anticipation of these experiments, we discuss here the 
quantum phases of dipolar fermions in harmonically confined 
optical lattices, paying particular attention to the emergence 
of superfluid phases that break time reversal symmetry spontaneously, 
as there are no confirmed analogues in condensed matter 
physics~\cite{footnote1}. 
By including the effects of screening, we show that 
quantum phases of dipolar fermions in harmonically confined optical 
lattices (two-dimensional geometry) can be approximately described 
by an extended Hubbard model for intermediate and high filling
factors, where only local on-site and a few neighbor 
interactions are required. For attractive local
and nearest neighbor interactions, we derive the phase diagram and
establish all accessible phases in the singlet channel. The most
important phases correspond to d-wave superfluidity and to 
superfluid phases that involve a superposition of $s$-wave and 
$d$-wave components of the order parameter and that break 
time reversal symmetry spontaneously. These phases naturally 
arise due to the non-local nature of dipolar interactions between 
heteronuclear molecules. In the broken time-reversal symmetry phases, 
spontaneous currents flow and can be detected using experimental 
techniques such as 
velocity sensitive Bragg spectroscopy~\cite{raman-2006, vale-2008}.

The bare Hamiltonian for dipolar fermions in optical lattices 
for a two-dimensional system ($xy$-plane) is
\begin{equation}
H_{BA}
= 
-t\sum_{\langle ij \rangle \sigma} c^{\dagger}_{i \sigma} c_{j \sigma}
+ 
U \sum_i n_{i \uparrow} n_{i \downarrow}
+ 
\sum_{i < j, \sigma \sigma^\prime} V_{ij} n_{i \sigma} n_{j \sigma^\prime},
\end{equation}
where $\langle ... \rangle$ indicates nearest neighbors, and 
$n_{i \sigma} = c^\dagger_{i \sigma} c_{i \sigma}$ is 
the local particle number operator. A complete derivation
of $H_{BA}$ is found in the supplemental material~\cite{supplemental-material}.
The on-site interaction $U = U_s + V_0$ contains two contributions. 
The first is from $s$-wave scattering 
$
U_s 
= 
(4\pi \hbar^2 a_s/m) 
\int d{\bf r}\vert w ({\bf r}) \vert^4
$ 
and the second is from the on-site dipole-dipole
interaction 
$
V_0 
= 
\int d{\bf k} 
w_F^2 ({\bf k}) 
V ({\bf k}),
$ 
where $a_s$ is the $s$-wave scattering length, $w({\bf r})$ is
the Wannier function, $w_F ({\bf k})$ and $V ({\bf k})$ are
the Fourier transforms of $\vert w ({\bf r}) \vert^2$ and of the
dipole-dipole potential $V_{ij}$, respectively. 
The long-range part of the dipole-dipole interaction is 
$
V_{ij} \approx D^2 
\left[
{\bf e}_i \cdot {\bf e}_j 
- 3 ( {\bf e}_i \cdot \hat {\bf r}_{ij} ) 
( {\bf e}_j \cdot \hat {\bf r}_{ij} ) 
\right]
\vert {\bf r}_i - {\bf r}_j \vert^{-3},
$ 
where $D$ is the magnitude of the dipole moments located at 
${\bf r}_i$ and ${\bf r}_j$,
${\bf e}_i$ is the direction of the dipole moment at ${\bf r}_i$, 
and $\hat {\bf r}_{ij}$ is the direction of the line connecting dipoles,
while ${\bf r}_i - {\bf r}_j$ is the distance between dipoles, 
with $r_j = (j_x a, j_y a)$ being the lattice vector, 
and $a$ being the lattice spacing.

The bare Hamiltonian is now transformed into an effective many-body
Hamiltonian which includes the effects of screening, as the 
interaction between two bare dipoles is renormalized (reduced) by
the non-local dielectric function 
$\epsilon_{NL} ({\boldsymbol \rho}, {\bf r}^{\prime\prime})$,
leading to an effective screened dipole-dipole interaction
$
V_{SC} ({\boldsymbol \rho}) 
= 
\sum_{{\bf r}^\prime}
V ({\bf r}^{\prime\prime}) 
\epsilon_{NL}^{-1} ({\boldsymbol \rho},{\bf r}^{\prime\prime}).
$
Although this effect is weak at small filling factors $(\nu < 0.05)$, 
it is substantial for filling factors $\nu > 0.1$ as the dielectric function 
becomes sufficiently large to reduce the range of the interaction to a 
few neighbors (See supplemental material~\cite{supplemental-material}).

Since we are interested in superfluid phases, we choose to tune the
experimental parameters to generate mostly attractive interactions.
For the off-site dipolar interactions, we choose to have all dipoles
aligned along the same direction $(\alpha, \phi)$ of an external
electric field, where $\alpha$ $(\phi)$ is the polar (azimuthal)
angle with respect to the $z$ ($x$) axis. The interaction becomes
${\widetilde V}_{ij} = V_x \delta_{\langle ij \rangle}$
(${\widetilde V}_{ij} = V_y \delta_{\langle ij \rangle}$) along the
$x$ ($y$) axis. Here, $V_x = D^2 (1 - 3 \sin^2 \alpha \cos^2
\phi)/[a^3 \epsilon_L (a)]$, $V_y = D^2 (1 - 3 \sin^2 \alpha \sin^2
\phi )/[ a^3 \epsilon_L (a) ]$, while $\delta_{\langle ij \rangle} =
1$ for nearest neighbors and zero otherwise. Here, $\epsilon_L ({\bf r})$
is the local dielecric function~\cite{supplemental-material}.
For the angles $\phi = \pm \pi/4,
\pm 3\pi/4$, the interactions are $V_x = V_y = V = D^2(1- 3 \sin^2
\alpha /2)/ [ a^3 \epsilon_L (a) ]$, and become negative when the
condition $\sin^2 \alpha > 2/3$ is satisfied.
For the on-site interaction, we choose to adjust the scattering 
length $a_s$ to produce $U = - \vert U \vert < 0$. We choose the electric
field to be parallel to the lattice plane with angles $\alpha = \pi/2,
\phi = \pi/4$. 

To study the physics discussed above, we use the effective
two-dimensional model hamiltonian
\begin{equation}
H = 
-t\sum_{\langle ij \rangle \sigma} c^{\dagger}_{i \sigma} c_{j \sigma}
- \vert U \vert \sum_i n_{i \uparrow} n_{i \downarrow}
- \vert V \vert \sum_{\langle ij \rangle \sigma \sigma^\prime}
n_{i \sigma} n_{j \sigma^\prime}
\end{equation}
on a square lattice, with first few neighbors hopping and interactions.
In order to establish the quantum phases as a function of
filling factor $\nu$, we start by constructing the partition function
$Z = \int {\cal D} c^\dagger {\cal D} c e^{S}$ for the action
\begin{equation}
\label{eqn:action}
S = \int_0^{\beta} d\tau
\left[
\sum_{i\sigma} c^{\dagger}_{i\sigma} (\tau)
(-\partial_\tau + \mu) c_{i\sigma} (\tau)
- H (c^\dagger, c)
\right].
\end{equation}

By symmetry, the singlet order parameters for superfluidity correspond to
local $s$-wave $\Delta_{sl}$, extended
$s$-wave $\Delta_{se}$ and $d$-wave $\Delta_{d}$ pairing~\cite{footnote2}. 
Upon a simple and standard integration of the fermionic degrees of freedom 
the action becomes
\begin{equation}
\label{eqn:action-order-parameter}
S = - \frac{N_s}{T}
\sum_{q, \alpha} \frac{\vert \Delta_{\alpha} (q) \vert^2}{V_{\alpha}}
+ Tr \ln \left( \frac {{\bf G}_{0}^{-1}}{T} - \frac{ {\bf V}}{T} \right)
+ \frac{\mu N_s}{T},
\end{equation}
where $\alpha = sl, se, d$; the interactions are
$V_{sl} = \vert U \vert$, $V_{se} = V_{d} = \vert V \vert$;
and the four-vector $q = (i\nu_n, {\bf q})$. 
The inverse free fermion propagator matrix is 
\begin{equation}
{\bf G}_{0}^{-1} (k, k^\prime) =
\left(
\begin{array}{cc}
 i \omega_n - \xi_{\bf k}  &   0 \\
0 & i \omega_n + \xi_{\bf k}
\end{array}
\right) \delta_{k, k^\prime}
\end{equation}
with kinetic energy $\xi_{\bf k} = \epsilon_{\bf k} - \mu$,
band dispersion 
$\epsilon_{\bf k} = - 2t \left[ \cos(k_x a) + \cos(k_y a) \right]$,
chemical potential $\mu$, four-vector $ k = (i\omega_n, {\bf k})$,
and unit cell length $a$. The bandwith of the dispersion is $w  = 8 t$.
The additional matrix appearing in Eq.~(\ref{eqn:action-order-parameter}) is
\begin{equation}
\label{eqn:interaction-matrix}
{\bf V} (k, k^\prime) =
\left(
\begin{array}{cc}
0 &  {\Delta}_{\alpha} (k - k^\prime) \\
{\Delta}_{\alpha}^{*} (-k + k^\prime) & 0
\end{array}
\right) \lambda_{\alpha}({\bf k}, {\bf k}^\prime ),
\end{equation}
where the Einstein summation over $\alpha$ is understood,
and $\lambda_{\alpha}({\bf k}, {\bf k}^\prime )$ are the
symmetry factors for the order parameters, which in the
limit of zero momentum pairing $({\bf k} = {\bf k}^\prime )$ become
$\lambda_{sl} ({\bf k}, {\bf k}) = 1$,
$\lambda_{se}({\bf k}, {\bf k}) = \cos(k_x a) + \cos(k_y a)$,
$\lambda_{d}({\bf k}, {\bf k}) = \cos(k_x a) - \cos(k_y a)$.

In terms of the quasiparticle
($\gamma = 2$) or quasihole ($\gamma = 1$) energies $E_{{\bf k},
\gamma} = (-)^\gamma \sqrt{ \xi_{\bf k}^2 + \vert {\Delta}_\alpha
\lambda_{\alpha} ({\bf k}) \vert^2 }, $ where the symmetry function
$\lambda_{\alpha} ({\bf k}) = \lambda_{\alpha} ({\bf k}, {\bf k})$,
the effective action becomes
\begin{equation}
\label{eqn:action-saddle-point}
S = - \frac{N_s}{\vert U \vert T} \vert {\Delta}_{sl} \vert^2
- \frac{N_s}{\vert V \vert T} ( \vert {\Delta}_{se} \vert^2 +
\vert {\Delta}_{d} \vert^2  )
+ S_2 + \frac{\mu N_s}{T}.
\end{equation}
Here, the second term in the action is $S_2 = \sum_{{\bf k},\gamma}
\ln \left[ 1 + \exp(-E_{ {\bf k},\gamma }/T ) \right]$. Notice that
there are three possible pure phases: local $s$-wave ($sl$); extended
$s$-wave ($se$) and $d$-wave ($d$). 
In addition, there are several possible binary
mixed phases $sl \pm se$, $sl \pm d$, and $se \pm d$, which do not
break time-reversal symmetry, and there are also
those that do, such as $sl \pm i se$, $sl \pm i d$, and $se \pm id$.
Lastly, several possible
ternary mixed phases involving all three symmetries $sl$, $se$ and
$d$ may also exist.

{\it General Case:} The order parameter equations can be obtained by
minimization of the action with respective to each order parameter.
By taking $\delta S/ \delta {\Delta}_{\alpha}^\ast = 0$, with $\alpha =
sl, se, d$,  we obtain
\begin{equation}
\label{eqn:order-parameter-general}
\Delta_{\alpha} = \frac{V_\alpha}{N_s} \sum_{\bf k} \frac{\tanh
(E_{{\bf k}, 2}/2T)} {2 E_{{\bf k}, 2} } \Lambda_{\alpha} ({\bf k}),
\end{equation}
with symmetry factors 
$
\Lambda_{\alpha} (\mathbf{k}) = \lambda_{\alpha} (\mathbf{k}) \big[
\Delta_{\alpha^\prime} \lambda_{\alpha^\prime} (\mathbf{k}) \big],
$
where repeated indices $\alpha^\prime$ indicate summation.

The number equation that fixes the chemical potential is obtained
through the thermodynamic relation $N = -\partial \Omega / \partial \mu$,
where $\Omega = - T \ln Z$ is the thermodynamic potential. In the
present approximation
$\Omega = -T S$, and the number equation reduces to
\begin{equation}
\label{eqn:number}
\nu  = \frac{1}{N_s} \sum_{\bf k}
\left[
1 - \frac{\xi_{\bf k}}{E_{{\bf k}, 2}}
\tanh ( E_{{\bf k}, 2}/2T )
\right],
\end{equation}
where $\nu = N/N_s$ is the filling factor.

\begin{figure}[htb]
\centering
\includegraphics[width = \linewidth]{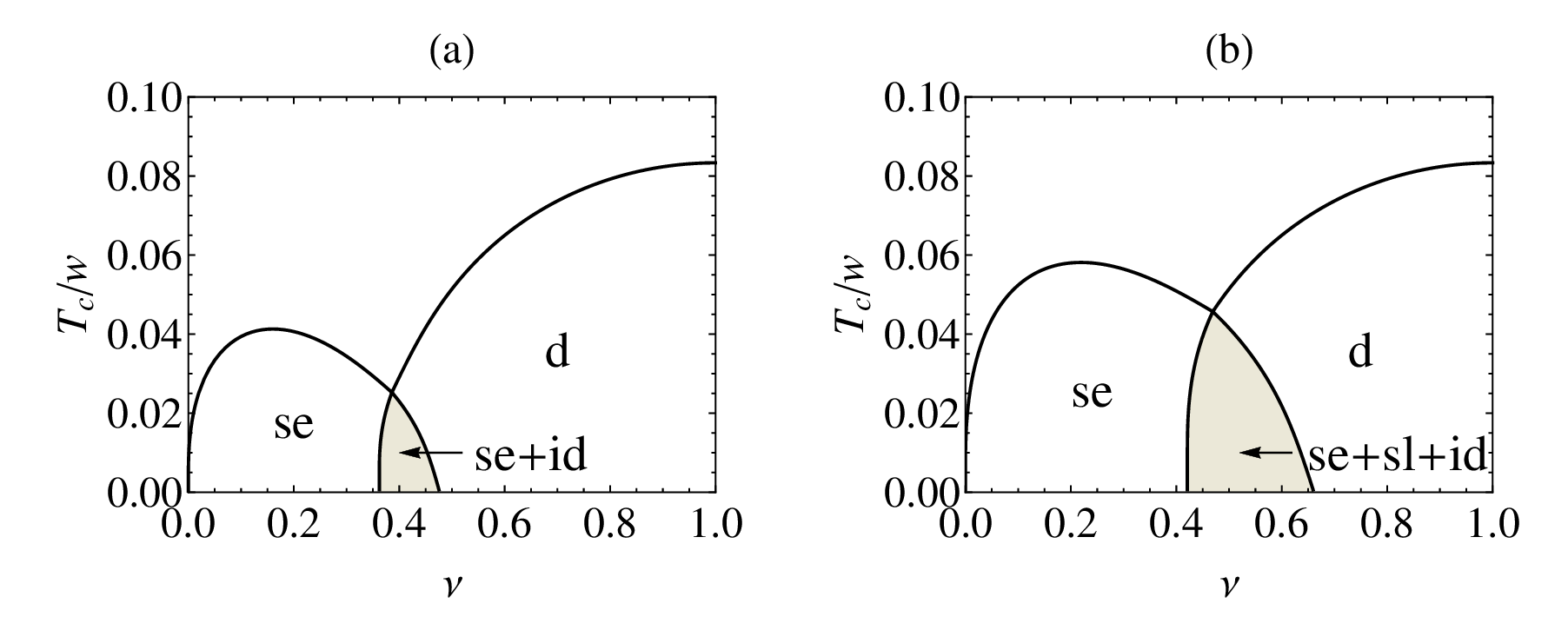} 
\caption{
\label{fig:one}
Critical temperature $T_c/w$ versus filling
factor $\nu$ at fixed interaction
$\vert U \vert/w = 0$ in (a)
or $\vert U \vert/w = 1/4$ in (b)
and $\vert V \vert/w = 3/8$. Notice the
tetracritical point where the normal and all superconducting
phases meet.
}
\end{figure}

Using the amplitude-phase representation, we write the order
parameters as ${\Delta}_{sl} = \vert {\Delta}_{sl} \vert
e^{i\phi_{sl}}$ for the local $s$-wave symmetry, ${\Delta}_{se} =
\vert {\Delta}_{se} \vert e^{i\phi_{se}}$ for the extended $s$-wave
symmetry, and ${\Delta}_d = \vert {\Delta}_d \vert e^{i\phi_d}$ for
the $d$-wave symmetry. The critical temperature can be obtained by
setting the order parameters $\Delta_{sl} = 0$, $\Delta_{se} = 0$,
and $\Delta_{d} = 0 $ in Eqs.~(\ref{eqn:order-parameter-general})
and~(\ref{eqn:number}).
In this case, the filling factor dependence of the critical
temperature $T_c (\nu)$ and the critical chemical potential $\mu_c
(\nu)$ can be obtained for pure $sl$-, $se$- and $d$-wave
symmetries. The solutions for $T_c (\nu)$ are shown in
Fig.~\ref{fig:one} for two cases $\vert U \vert/w = 0$ 
and $\vert V \vert/w = 3/8$,  
as well as $\vert U \vert/w = 1/4$ and $\vert V \vert /w = 3/8$,
where the corresponding superfluid phases are also indicated. 
The phase diagram obtained within the saddle point 
approximation is very accurate provided that
both $\vert V \vert^2/w^2 \ll 1$ and $\vert U \vert^2/w^2 \ll 1$,
but it is only semi-quantitative when 
$\vert V \vert^2/w^2 \lesssim 1$ or $\vert U \vert^2/w^2 \lesssim 1$.
The phase diagram is symmetric about $\nu = 1$, since the Helmholtz free
energy $F = \Omega + \mu N$ is invariant under the global
particle-hole transformation $\mu \to - \mu$ and $\nu \to 2 - \nu$.
Notice that $s$-wave phases are favored at
lower filling factors, while the $d$-wave phase is favored near
half-filling, this is directly correlated with the higher effective
density of states in this vicinity. The
time-reversal-symmetry-breaking phases occur at filling factors
between the $s$-wave and $d$-wave phases. Generally, when 
$\vert U \vert/w = 0$, the only accessible phases are $se$-, $d$- and $(se
\pm id)$-wave, and when $\vert U \vert/w \ne 0$ the only accessible
phases are $(sl + se)$-, $d$ and $(sl + se \pm id)$-wave.
%

%
%

Given that on-site interactions can be experimentally controlled, we
focus our discussion at $\vert U \vert/w = 0$, which already contains the
essential physics of superfluid phases that spontaneously break time
reversal symmetry and have a $d$-wave component. The Ginzburg-Landau
theory near $T_c$ is obtained by expanding the action of
Eq.~(\ref{eqn:action-saddle-point}) in terms of the order parameters
$\Delta_{se}$, $\Delta_d$ and their complex conjugates. From the
thermodynamic potential $\Omega = - TS$, we can calculate the
Helmoltz Free energy $F = \Omega + \mu N$. The free energy per site
${\cal F} = F/N_s$ takes the simple form
\begin{equation}
\label{eqn:free-energy-near-critical-temperature}
\begin{array}{c}
{\cal F} =
a_{se} \vert \Delta_{se} \vert^2 +
a_d \vert \Delta_d \vert^2 +
b_{se} \vert \Delta_{se} \vert^4 +
b_d \vert \Delta_d \vert^4 + \\
2 b_{sd} \left[ 1 + \frac {1}{2} \cos (2 \delta \phi) \right]
\vert \Delta_{se} \vert^2  \vert \Delta_d \vert^2
+ \mu (\nu - 1),
\end{array}
\end{equation}
when the thermodynamic potential $\Omega$ is expanded to fourth order 
in the order parameters using the action $S$ defined 
in Eq.~(\ref{eqn:order-parameter-general}).
The coefficients $a$ and $b$ depend explicitly on the parameters of the
model used. In the present case the possible phases
$se \pm d$ are not accessible, and a tetracritical point exists 
where the normal and superconducting phases with $se$, $d$ and 
$se \pm id$ symmetries meet.
In addition, the free energy depends only on $2 \delta \phi$ and
does not distinguish between the phases $se + id$ and $se - id$,
which are thus degenerate.
In the $se \pm id$ phases, time-reversal symmetry is broken
but not chirality.

%
%

\begin{figure}[htb]
\centering
\includegraphics[width = \linewidth]{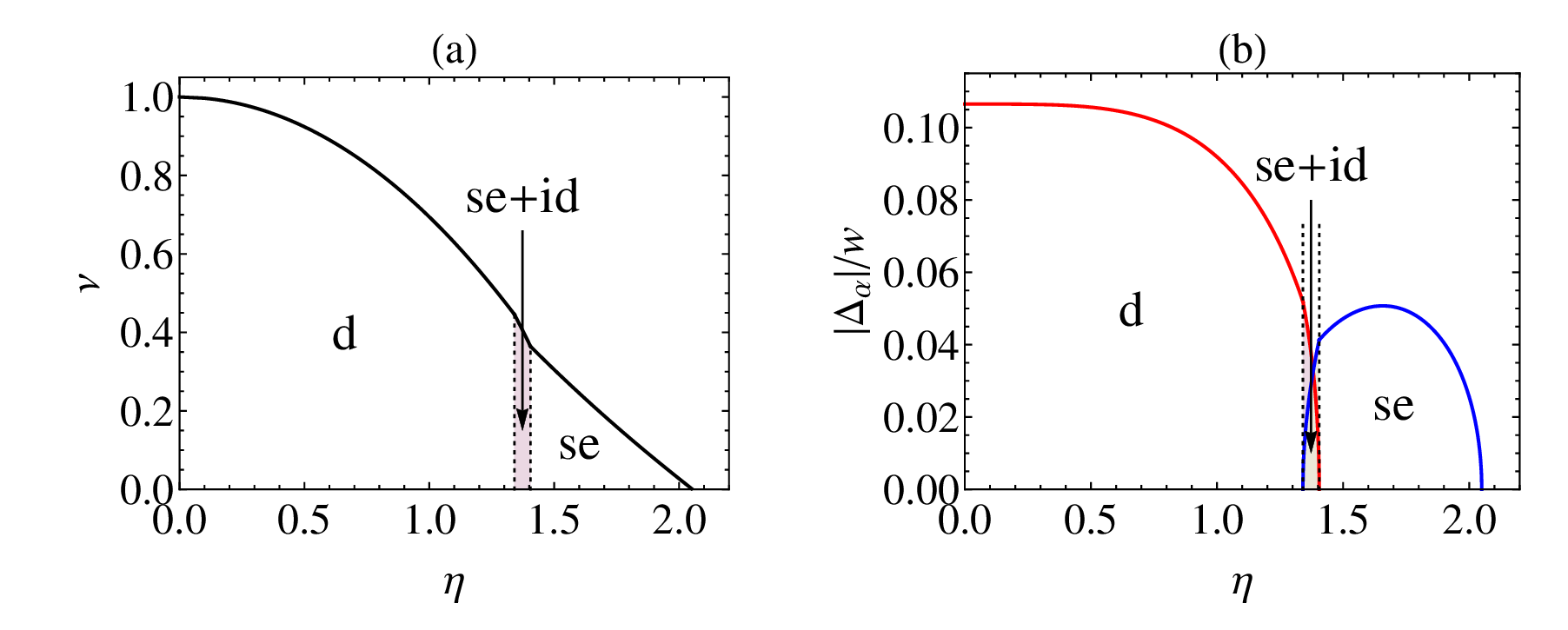}
\caption{
\label{fig:two}
Spatially resolved filling factors $\nu$ in (a) and superfluid
order parameters $\Delta_\alpha$ $(\alpha = se, d)$ in (b)
as a function of $\eta = \left[ w/(8 \epsilon_h)\right]^{1/2}(r/a)$,
for $\nu (0) = 1$ (half-filling)
at center of trap and parameters
$\vert U \vert/w = 0$, $\vert V \vert/w = 3/8$, and $T/w = 0.0125$.
}
\end{figure}

{\it Harmonic Trap:} The essential effect of an underlying harmonic
trap $V_h ({\bf r})  = k r^2/2$ is to allow for the emergence of
non-uniform solutions. In particular, the harmonically confining
potential allows for the existence of all accessible phases $se$-,
$d$- and $(se \pm id)$-wave for $\vert U \vert/w = 0$ and $sl + se$,
$d$- and $(sl + se \pm id)$-wave for $\vert U \vert/w \ne 0$. Within
the local density approximation, we solve the order parameter
Eq.~(\ref{eqn:order-parameter-general}) 
and number Eq.~(\ref{eqn:number}) with
$\mu \to \mu - V_h ({\bf r})$. We obtain the profiles of the filling
factor and order parameters, shown in Fig.~\ref{fig:two}, as a
function of dimensionless position from the center of the trap 
$
\eta = 
\left[ 
w/(8\epsilon_h)
\right]^{1/2}
(r/a),
$
where $\epsilon_h = k a^2/2$, and for 
parameters $\vert U \vert/w = 0$, $\vert V \vert/w = 3/8$, 
$T/w = 0.0125$, assuming that $\nu = 1$ (half-filling) 
at the center of the trap. Notice that as the filling factor 
decreases from the center of
the trap to its edge, all accessible phases emerge: $d$-wave
superfluid at the center of the trap, followed sequentially by
regions of $(se \pm id)$- an $se$-wave superfluid, and the normal
state. Similarly, in the case of $\vert U \vert/w \ne 0$ at low
temperatures and assuming that $\nu = 1$ at the center of the trap,
the sequence of phases from the center of the trap is $d$-, $(sl +
se \pm id)$-, $(sl + se)$-wave superfluid followed by a normal
region at the edge. The interesting qualitative aspect here is the
emergence of regions where time-reversal symmetry is spontaneously
broken: $(se \pm id)$ for $\vert U \vert/w = 0$ and $sl + se \pm id$
for $\vert U \vert/w \ne 0$. This is very important in a very broad
sense, because there are no confirmed examples in condensed matter
physics of superfluids that spontaneously break-time-reversal
symmetry.~\cite{footnote1}
%

%
%

{\it Spontaneous Currents:}
In order to keep the discussion simple, we continue to focus on the case of
$\vert U \vert/w = 0$, and
discuss the spontaneous current flow in the shell corresponding to the
$(se \pm id)$-wave superfluid.
Consider for example that
either the $se + id$ phase or the $se - id$ phase is realized
in the example of Fig.~\ref{fig:two}.
Given that either chiral phase
spontaneously break time-reversal symmetry, it is expected that within
the boundaries of the $se + id$ $(se - id)$ phase spontaneous
currents circulate clockwise (counter-clockwise) near the outer boundary, and
counter-clockwise (clockwise) near the inner boundary. To visualize 
the spontaneously generated currents, we perform a long-wavelength 
expansion of the action in Eq.~(\ref{eqn:action-order-parameter}), 
which leads to the effective Free energy density
$
{\cal F}_{\rm eff}
=
{\cal F}_{\rm di}
+
{\cal F}_{\rm nd}
+
{\cal F}_{\rm h}
+
{\cal F}.
$
The first term is
\begin{equation}
{\cal F}_{\rm di}
=
\nabla
\Delta_{se}^*
\frac{c_{se,se}}{2m}
\nabla
\Delta_{se}
+
\nabla
\Delta_{d}^*
\frac{c_{d,d}}{2m}
\nabla
\Delta_{d},
\nonumber
\end{equation}
the second term is non-diagonal in the indices $se$ and $d$
\begin{equation}
{\cal F}_{\rm nd}
=
\left[
\partial_x
\Delta_{se}^*
\frac{c_{se,d}}{2m}
\partial_x
\Delta_{d}
-
\partial_y
\Delta_{se}^*
\frac{c_{se,d}}{2m}
\partial_y
\Delta_{d}
+
C. C.
\right],
\nonumber
\end{equation}
the third term is
$
{\cal F}_{\rm h} =
\gamma_{se} V_h ({\bf r}) \vert \Delta_{se} \vert^2
+
\gamma_{d} V_h ({\bf r}) \vert \Delta_{d} \vert^2,
$
while the last term ${\cal F}$ is given in
Eq.~(\ref{eqn:free-energy-near-critical-temperature}).
Adding a current source term
$-i\partial_m - a_m$
and considering the phase difference $\delta \phi = \phi_d - \phi_{se} = \pm \pi/2$,
we obtain within the $se \pm id$ phase the particle current density
$
J_i
=
J_{i, \phi}
+
J_{i, \vert \Delta \vert},
$
in Cartesian representation $(i = x, y)$.
Here,
$
J_{i, \phi}
=
\frac{2}{m}
\left[
\vert
\Delta_{d}
\vert^2
c_{d,d}
+
\vert
\Delta_{se}
\vert^2
c_{se,se}
\right]
\partial_i \phi_{d}
$
is a phase-related contribution and 
$
J_{i,\vert \Delta
\vert} = \frac{2}{m} \chi \left[ \vert \Delta_{se} \vert c_{i, se,d}
\partial_i \vert \Delta_{d} \vert
-
\vert \Delta_{d} \vert
c_{i, se, d}
\partial_i \vert \Delta_{se} \vert
\right] $ 
is an amplitude-related component, where $\chi = \sin
(\delta\phi) = \pm 1$ is the chirality of the $se \pm id$ phase. In
addition, the coefficients $c_{i, se, d}$ satisfy the relation
$c_{x, se, d} = - c_{y, se, d} = c_{se, d}$. 
Given the existence of the harmonic potential, we transform the
currents to polar coordinates $(r, \theta)$, and require
the radial current $J_r$ to vanish 
$( J_r = {\bf \hat r} \cdot {\bf J} = 0 )$ at the
boundaries between the $se \pm i d$ and $se$ occurring at $r =
R_{se}$ and at the boundaries between $se \pm id$ and $d$ phases
occurring at $r = R_{d}$. At these boundaries spontaneous currents
flow only within the $se \pm id$ phase limits, since these are the
only phases that break spontaneously time-reversal symmetry. Under
these conditions, non-trivial solutions for 
$
\phi_d 
= 
\chi
\left[ \pi/2 + f(r) \theta \right]
$ 
and 
$
\phi_{se} 
= 
\chi f(r) \theta
$
are possible
with 
boundary conditions
$f(r = R_{se}) = +1,$, 
$f(r = R_{d}) = -1,$ 
and
$
df(r)/dr\vert_{R_{se}} 
= 
df(r)/dr\vert_{R_{d}} 
= 
0.
$
The spontaneous currents at the interface boundaries are tangential,
having the forms $ J_{\theta} ( r = R_{se} ) \approx (2\chi/m) \vert
\Delta_d \vert^2 c_{d,d}/R_{se} $ and 
$ 
J_{\theta} ( r = R_{d} ) 
\approx
-(2\chi/m) 
\vert \Delta_{se} 
\vert^2 c_{se,se}/R_{d}, 
$
which can be detected via Bragg spectroscopy as discussed next.

%
%

\begin{figure}[htb]
\centering
\begin{tabular}{cc}
\includegraphics[width = 0.5\linewidth]{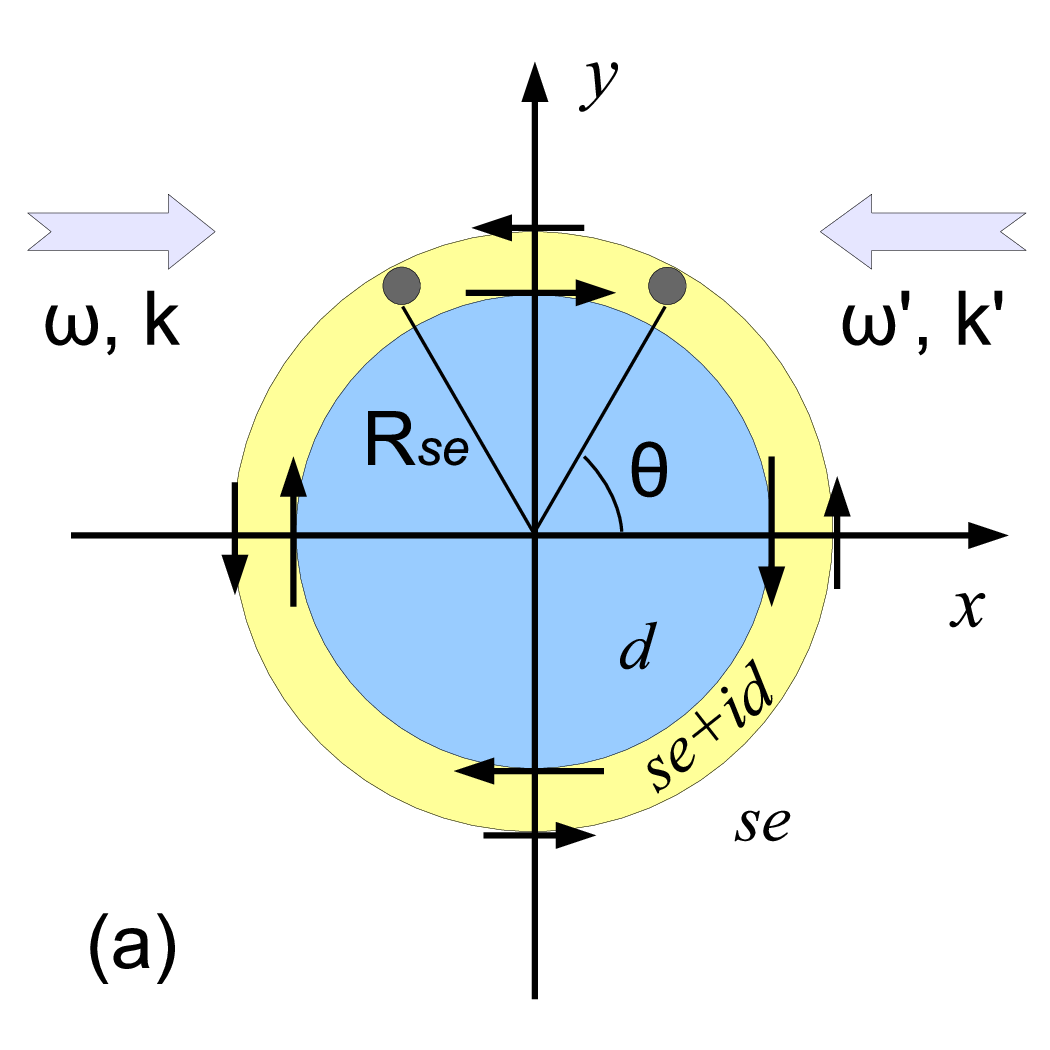} \quad &
\includegraphics[width = 0.5\linewidth]{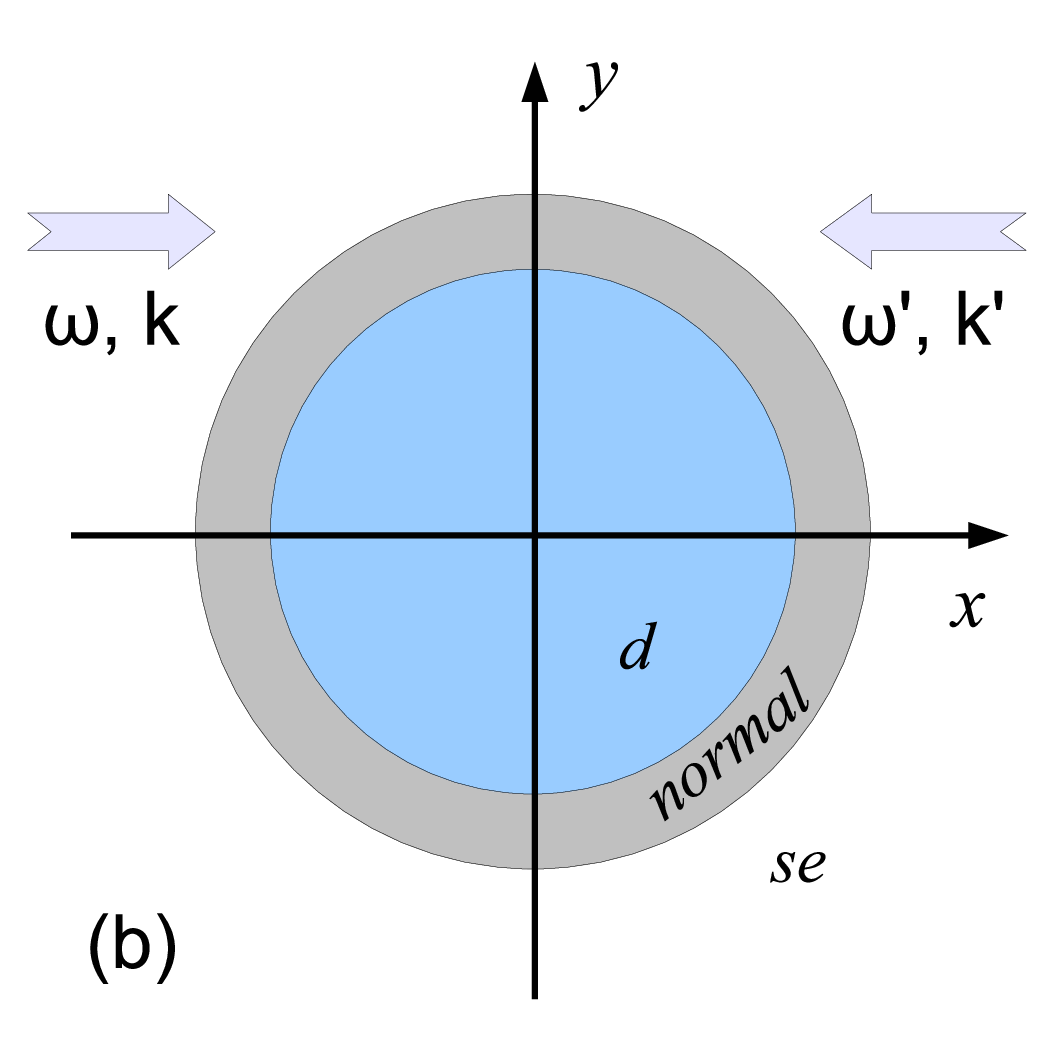}
\end{tabular}
\caption{
\label{fig:three}
Schematic plots of a velocity sensitive Bragg spectroscopy scheme
to detect circulating currents of superfluid regions that
break time reversal symmetry spontaneously.
The case illustrated corresponds to $\vert U \vert/w =0$
with filling factor $\nu = 1$ at the center of the trap.
}
\end{figure}

{\it Detection of time-reversal-symmetry-breaking:}
A detection scheme of spontaneous currents using velocity
sensitive Bragg spectroscopy~\cite{raman-2006, vale-2008}
is shown in Fig.~\ref{fig:three} with right- (left-) going beam of frequency
$\omega$ $(\omega^\prime)$ and linear momentum ${\bf k}$ $({\bf k}^{\prime})$.
In Fig.~\ref{fig:three}a, circulating currents are shown at the
boundaries of the region for $(se + id)$ superfluidity,
due to spontaneous breaking of time reversal symmetry at lower temperatures.
The case of a normal region (higher temperatures),
where no spontaneous currents exist, is shown
in Fig.~\ref{fig:three}b for comparison. In Fig.~\ref{fig:three}a, 
the two dark spots making angles $\theta$ and 
$\pi - \theta$ with the horizontal satisfy the Bragg conditions due to 
the Doppler shift creating by circulating currents in the $(se + id)$ region.
Such Bragg spots are inexistent when there are no-circulating currents present,
as is the case of the normal-state shown in Fig.~\ref{fig:three}b. 
%

%
%

{\it Chemical and Collisional Stability:} The chemical and collisional 
stability of candidate molecules is a very important issue. Currently is
known that fermionic molecules such as ${\rm LiCs}$ and 
${\rm KRb}$ are not chemically stable~\cite{hutson-2010}, 
and tend to decay through collisions into ${\rm Li}_2$ and 
${\rm Cs}_2$ or ${\rm K}_2$ and ${\rm Rb}_2$, respectively, and 
thus are not ideal candidates for the effects proposed here. 
However, fermionic ${\rm NaK}$ is chemically stable, 
has a hyperfine structure, and the hyperfine 
states in electronic-roto-vibrational ground state may have sufficiently 
long lifetimes~\cite{NaK-molecules-2013}, thus making it
an ideal candidate for the effects proposed 
here~\cite{supplemental-material}.

%
%

{\it Conclusions:} We discussed
screened dipolar fermions in harmonically confined optical lattices
modeled by an extended attractive Hubbard model,
where both interactions and filling factors can be controlled.
We had in mind particularly the fermionic dipolar molecule 
$^{23}{\rm Na} ^{40}{\rm K}$ which is chemically stable in its 
electronic-roto-vibrational ground state, but presents
a hyperfine structure allowing for the creation of two-mixed spin states.
We established the superfluid phases in the singlet channel and
indicated that accessible phases have not only pure $s$-wave or 
$d$-wave characters, but also mixed $(s \pm id)$-wave character 
which breaks time
reversal symmetry spontaneously.
We calculated the spatially-dependent profiles of filling factor 
and order-parameter
for various superfluid phases, and proposed a Bragg spectroscopy
experiment to detect the time-reversal symmetry breaking phase, which contains
spontaneously circulating supercurrents.

\acknowledgements{We thank ARO (Grant No. W911NF-09-1-0220) for support.}

\pagebreak
\widetext
\vspace{0.2in}
\begin{center}
\textbf{\large Supplemental Materials: Superfluid Phases of Dipolar Fermions in Harmonically Trapped Optical Lattices} 
\end{center}
\setcounter{equation}{0}
\setcounter{figure}{0}
\setcounter{table}{0}
\setcounter{page}{1}
\makeatletter
\renewcommand{\theequation}{S\arabic{equation}}
\renewcommand{\thefigure}{S\arabic{figure}}
\renewcommand{\bibnumfmt}[1]{[S#1]}
\renewcommand{\citenumfont}[1]{S#1}

\vspace{0.2in}

In this supplemental material, we provide details of the
construction of the effective lattice hamiltonian in the presence of 
long-range dipolar interactions, the effects of screening
and a discussion of collisional properties of dipolar molecules.

\vspace{0.1in}

To describe the superfluid phases of ultracold dipolar fermions 
in optical lattices for two-dimensional systems, first 
we derive in this supplemental material the lattice Hamitonian used
to obtain distinct pairing symmetries that naturally arise in 
the singlet channel: local $s$-wave $(sl)$, 
extended $s$-wave $(se)$, $d$-wave $(d)$
or time-reversal-symmetry breaking $(sl + se \pm id)$-wave.
Second, we discuss the screening effects of the dipolar interactions 
within the random phase approximation.
Finally, we comment on the effects of chemical
and collisional stability of dipolar molecules, and
suggest that fermionic ${\rm NaK}$ molecules are 
potentially a very good candidate for the emergence 
of the superfluid phases discussed in the main text.

\subsection{Effective Lattice Hamiltonian}
To obtain the effective lattice Hamiltonian described 
in this manuscript, we start with dipolar molecules confined
to move in two-dimensions and described by the following
continuum Hamiltonian 
\begin{equation}
H_C =  H_{SP} + H_{SR} + H_{LR},
\end{equation}
where the first term represents the single particle
energy 
\begin{equation}
H_{SP}
=
\int d{\bf r} 
\psi_\sigma^\dagger ({\bf r})
\left[
{\hat K}
+
V_P ({\bf r})
\right]
\psi_\sigma ({\bf r}),
\end{equation}
where ${\hat K} = -\nabla^2/(2m)$ is
the kinetic energy operator $(\hbar = 1)$ and
$V_P ({\bf r})$ is a periodic potential that
produces a square lattice pattern. 
The second term represents the short-ranged (local) attractive
contact interaction
\begin{equation}
H_{SR}
=
-
g
\int d {\bf r}
\psi_\uparrow^\dagger ({\bf r})
\psi_\downarrow^\dagger ({\bf r})
\psi_\downarrow ({\bf r})
\psi_\uparrow ({\bf r})
\end{equation}
and the last term represents
the long-range interactions
\begin{equation}
H_{LR}
=
\frac{1}{2}
\int d{\bf r} d{\bf r}^\prime
V_{LR} ({\bf r},{\bf r}^\prime)
\psi_\sigma^\dagger ({\bf r})
\psi_{\sigma^\prime}^\dagger ({\bf r}^\prime)
\psi_{\sigma^\prime}({\bf r}^\prime)
\psi_\sigma ({\bf r}).
\end{equation}
Here, the long-range interaction is described by the term
%
$$
V_{LR} ({\bf r},{\bf r}^\prime) 
= 
\frac{1}{4\pi \epsilon_0}
\left[
\frac{2 q^2}{\vert {\bf r} - {\bf r}^\prime \vert}
-
\frac{q^2}{\vert {\bf r} - {\bf r}^\prime  + {\bf d} \vert}
-
\frac{q^2}{\vert {\bf r} - {\bf r}^\prime  - {\bf d} \vert}
\right],
$$
%
which represents the Coulombic interaction between dipoles with 
effective charges $+q$ and $-q$, separated by the characteristic
distance $\vert {\bf d} \vert$. All the dipoles are assumed to be 
aligned along the same direction of a large external 
electric field ${\bf E}$, such that ${\bf d} \parallel {\bf E}$. 
The position vectors ${\bf r}$ and ${\bf r}^\prime$ reside on the $xy$ plane. 

Noticing that the minima of the optical lattice potential $V_P ({\bf r})$ 
define the lattice sites, we can write the lattice-fermion creation
operators as 
$
\psi_{\sigma}^{\dagger} ({\bf r}) 
= 
\sum_{i} 
\varphi_{i\sigma}^* ({\bf r})
c_{i \sigma}^\dagger
$
and the anihilation operators as 
$
\psi_{\sigma}({\bf r}) 
= 
\sum_{i} 
\varphi_{i\sigma} ({\bf r})
c_{i \sigma}.
$
Here, the Wannier functions $\varphi_{i\sigma} ({\bf r})$ obey the
orthonormality condition 
$
\int d{\bf r} 
\varphi_{i \sigma}^* ({\bf r})
\varphi_{j \sigma^\prime} ({\bf r})
= 
\delta_{ij} \delta_{\sigma \sigma^\prime}.
$
In the local Wannier basis, each contribution to the Hamiltonian becomes
\begin{equation}
H_{SP}
=
-
\sum_{i \sigma}
\epsilon_i 
c_{i \sigma}^\dagger c_{j \sigma}
- \sum_{i \ne j \sigma}
t_{ij \sigma}
c_{i \sigma}^\dagger c_{j \sigma}, 
\end{equation}
where the total on-site (local) energy is
$
\epsilon_i 
= 
-\int d{\bf r}
\varphi_{i \sigma}^* ({\bf r}) 
\left[
-\nabla^2/(2m)
+ 
V_P ({\bf r})  
\right]
\varphi_{i \sigma} ({\bf r})
$
is independent of the site due to translational
invariance 
$
(\epsilon_i 
= 
\epsilon_0),
$ 
and the hopping matrix elements are
$
t_{ij\sigma} 
=
\int d{\bf r} 
\varphi_{i \sigma}^* ({\bf r}) 
\left[
-\nabla^2/(2m)
+ 
V_P ({\bf r})  
\right]
\varphi_{j \sigma} ({\bf r}).
$
We take $\epsilon_0$ to be our reference energy
and set the on-site energy $\epsilon_0 = 0$. 

The short-range interaction is written as
\begin{equation}
H_{SR}
=
\sum_{ijk\ell}
U_{ijk\ell} 
c^\dagger_{i \uparrow} c^\dagger_{j \downarrow}
c_{k \downarrow} c_{\ell \uparrow},
\end{equation}
where 
$
U_{ijk\ell} 
=
-g 
\int {\bf d r}
\varphi_{i \uparrow}^* ({\bf r}) 
\varphi_{j \downarrow}^* ({\bf r}) 
\varphi_{k \downarrow} ({\bf r}) 
\varphi_{\ell \uparrow} ({\bf r}). 
$
Due to the orthonormality of the Wannier functions, the main contribution 
to $H_{SR}$ comes from $U_{iiii} = U_s$, while otherwise 
$U_{ijk\ell} = 0$. This implies that the contribution from short-ranged 
s-wave interactions is described by the on-site interaction 
\begin{equation}
H_{SR}
=
\sum_{i}
U_s
c^\dagger_{i \uparrow} c^\dagger_{i \downarrow}
c_{i \downarrow} c_{i \uparrow},
\end{equation}
with $U_s = -g \int d{\bf r} \vert w ({\bf r}) \vert^4$,
with $g = 4\pi \hbar^2 a_s/m$, and 
where we used the simplification 
$
\varphi_{i \uparrow} ({\bf r}) 
= 
\varphi_{i \downarrow} ({\bf r})
= 
w ({\bf r}).
$

Similarly the long-range part of the Hamiltonian
can be written as
\begin{equation}
H_{LR}
=
\sum_{ijk\ell}
V_{ijk\ell}^{\sigma \sigma^\prime}
c^\dagger_{i \sigma} c^\dagger_{j \sigma^\prime}
c_{k \sigma^\prime} c_{\ell \sigma},
\end{equation}
where the general matrix element has the form
$$
V_{ijk\ell}^{\sigma \sigma^\prime}
=
\frac{1}{2}
\int d {\bf r} d {\bf r}^\prime
V_{LR} ( {\bf r}, {\bf r}^\prime )
\varphi_{i\sigma}^* ({\bf r})
\varphi_{j\sigma^\prime}^* ({\bf r}^\prime)
\varphi_{k\sigma^\prime} ({\bf r}^\prime)
\varphi_{\ell\sigma} ({\bf r}).
$$
Due to the orthonormality of the Wannier functions, 
the dominant contributions are those corresponding to 
$
V_{iiii}^{\sigma \sigma^\prime} 
= 
(1/2) \int d{\bf r} d{\bf r}^\prime 
V_{LR} ( {\bf r}, {\bf r}^\prime )
\vert \varphi_{i \sigma} ({\bf r}) \vert^2
\vert \varphi_{i \sigma^\prime} ({\bf r}^\prime) \vert^2
$
which is effectively spin-independent,
since the simplification 
$
\varphi_{i \uparrow} ({\bf r}) 
= 
\varphi_{i \downarrow} ({\bf r})
= 
w ({\bf r})
$
holds in the present case,
leading to 
$
V_{iiii}^{\sigma \sigma^\prime} 
= 
V_0/2
=
(1/2) \int d{\bf r} d{\bf r}^\prime 
V_{LR} ( {\bf r}, {\bf r}^\prime )
\vert w ({\bf r}) \vert^2
\vert w  ({\bf r}^\prime) \vert^2.
$
The on-site contribution of the long-range
interactions can be written in Fourier
space as 
$
V_0 
= 
\int d{\bf k} V_{LR} ({\bf k})
\vert w_F ({\bf k}) \vert^2,
$
where $V_{LR} ({\bf k})$ is the Fourier transform
of 
$
V_{LR} ( {\bf r}, {\bf r}^\prime ) 
=
V_{LR} ( {\bf r} - {\bf r}^\prime ),$ 
and
$w_F ({\bf k})$ is the Fourier transform
of $\vert w ({\bf r})\vert^2$.

The last contribution to the lattice Hamiltonian is
$
V_{ijji}^{\sigma \sigma^\prime} 
= 
(1/2) \int d{\bf r} d{\bf r}^\prime 
V_{LR} ( {\bf r}, {\bf r}^\prime )
\vert \varphi_{i \sigma}({\bf r}) \vert^2
\vert \varphi_{j \sigma^\prime} ({\bf r}^\prime) \vert^2,
$
which is also effectively spin-independent, 
and corresponds to a density-density 
interaction with 
$
V_{ijji}^{\sigma \sigma^\prime} 
\approx 
V_{LR} ({\bf r}_i - {\bf r}_j)
= V_{LR} (i,j).
$
All the other terms from the tensor $V_{ijk\ell}^{\sigma \sigma^\prime}$ 
are comparatively small due to the orthonormality of the Wannier functions,
leading to the simplified expression
\begin{equation}
H_{LR}
=
\sum_{i} {V_0} 
c^\dagger_{i \uparrow} c^\dagger_{i \downarrow}
c_{i \downarrow} c_{i \uparrow}
+
\sum_{i < j, \sigma \sigma^\prime} 
V_{LR} (i,j)
c^\dagger_{i \sigma} c^\dagger_{j \sigma^\prime}
c_{j \sigma^\prime} c_{i \sigma}.
\end{equation}
The second term becomes the dipolar interaction
at large separations $\vert {\bf r}_i - {\bf r}_j \vert \gg d.$
The final lattice Hamiltonian is then the sum of the three contributions
$
H_{LT} 
=
H_K + H_L + H_{LR}
$
leading to the bare lattice Hamiltonian
\begin{equation}
H_{BA}
= 
-t\sum_{\langle ij \rangle \sigma} c^{\dagger}_{i \sigma} c_{j \sigma}
+ 
U \sum_i n_{i \uparrow} n_{i \downarrow}
+ 
\sum_{i < j, \sigma \sigma^\prime} V_{ij} n_{i \sigma} n_{j \sigma^\prime},
\end{equation}
where the local interaction is $U = U_s + V_0$ and the non-local
interaction is $V_{ij} = V_{LR}(i,j).$ However, like the Coulomb 
interaction in solid state crystals, the long-ranged interaction 
of dipolar nature can also be screened, as discussed next.

\subsection{Screening effects}
It is well stablished that screening effects are very important
at sufficiently large densities for electronic materials
which interact via long-ranged Coulomb forces~\cite{mahan-book}. 
In such systems, the effective interactions between electrons 
can be reduced to purely on-site or to on-site and nearest neighbors. 
Screening effects serve as the basis for the justification of 
simplified lattice models in condensed matter physics, such 
as the Hubbard model, where only on-site interactions 
are considered, or the extended Hubbard model with on-site and 
nearest neighbor interactions.

In the case of dipolar interactions, screening can also be important if 
the density of dipoles is sufficiently large. 
If in real space the bare interactions 
$V_{BA} ({\boldsymbol \rho})$ depend only on the separation 
${\boldsymbol \rho} = {\bf r} - {\bf r}^\prime$
between particles located at positions ${\bf r}$ and ${\bf r}^\prime$, then in
momentum space the screened interactions, in their simplest description,
can be expressed as a ladder sum of repeated interaction 
events~\cite{mahan-book} leading to
\begin{equation}
V_{SC} ({\bf q}) 
=
\frac{V_{BA}({\bf q})}{ 1 - V_{BA} ({\bf q}) P ({\bf q}) }, 
\end{equation}
where $P ({\bf q})$ is the zero-frequency polarization function for fermions,
producing static screening of the bare interaction.
At the first level of approximation $P ({\bf q})$ can be replaced
by the non-interacting polarization at zero frequency 
$$
P_0 ({\bf q})
= 
\sum_{\bf k}
\frac{n_F (\xi_{{\bf k} + {\bf q}}) - n_F (\xi_{{\bf k}})}   
{\xi_{{\bf k} + {\bf q}} - \xi_{\bf k} }. 
$$
The standard approach used here is called the static random 
phase approximation (RPA) for screening~\cite{mahan-book}.

The expression for the screened interaction in momentum space becomes
$
V_{SC} ({\bf q}) 
= 
V_{BA} ({\bf q})
/
\epsilon ({\bf q}) 
= 
V_{BA} \chi ({\bf q}),
$
where $\epsilon ({\bf q}) =  1 - V_{BA} ({\bf q}) P ({\bf q})$ is the 
dielectric function and $\chi ({\bf q})$ is the electric permittivity.
The screening interaction potential in real space becomes
\begin{equation}
V_{SC} ({\boldsymbol \rho})
=
\frac{1}{V_D} 
\int d{\bf r}^{\prime\prime} 
V_{BA} ({\bf r}^{\prime \prime})
\epsilon_{NL}^{-1} ( {\boldsymbol \rho}, {\bf r}^{\prime\prime} )
\end{equation}
where the non-local screening function is 
$
\epsilon_{NL} ( {\boldsymbol \rho}, {\bf r}^{\prime\prime} )
= 
\chi^{-1} ({\boldsymbol \rho} - {\bf r}^{\prime\prime}).
$
The screened interaction can be finaly written as
\begin{equation}
V_{SC} ({\boldsymbol \rho})
=
V_{BA} ({\boldsymbol \rho})/\epsilon_L ({\boldsymbol \rho}),
\end{equation}
where the local screening function is defined to be
$
\epsilon_L ({\boldsymbol \rho})
= 
V_{BA} ({\boldsymbol \rho})
/
\left[
V_D^{-1}
\int d {\bf r}^{\prime \prime}
V_{BA} ({\bf r}^{\prime \prime}) 
\chi ( {\boldsymbol \rho} - {\bf r}^{\prime\prime} )
\right].
$

In solids, for electrons interacting only via Coulomb repulsion, 
it is well established that screening plays a very important role
and leads to an effective lattice Hamiltonian that includes only 
local Coulomb (on-site Hubbard) interactions and nearest neighbors 
screened Coulomb (extended Hubbard) interactions, as 
can be inferred from standard many-body textbooks~\cite{mahan-book}. 
Such effective Hamiltonian is meant to describe quite accuratelly 
electrons interacing via Coulomb forces in crystal structures
at nearly any filling factor of the electronic band, with the sole
exception of very low filling factor, where screening is not effective
and the long-ranged nature of the Coulomb forces needs to be taken
into account. We performed a similar analysis here for long-ranged 
dipolar interactions using the random phase approximation 
described above, and find that the long-range dipolar interactions 
in an optical lattice are weakly screened for filling factors 
$\nu < 0.05$, but they are strongly screened beyond $\nu > 0.1$. This
indicates that for filling factors larger $\nu = 0.1$, we need to 
consider at most interaction between the first few neighbors, which 
thus justifies the use of the screened Hamiltonian that produces the phase
diagrams shown in Figs. 2 and 3 of the main text.

\subsection{Chemical and Collisional Stability}
 
Much of the experimental effort involving dipolar molecules
has been devoted to heteronuclear dimers 
consisting of alkali atoms~\cite{KRb-molecules-1,KRb-molecules-2, 
LiCs-molecules, RbCs-molecules}. It is known experimentally that fermionic 
${\rm ^{40}K ^{87}Rb}$ molecules are 
chemically unstable~\cite{Jin-Ye-2011} towards the formation 
of dimers ${\rm K_2}$ and ${\rm Rb_2}$. In addition, theoretical work 
have shown 
that all heteronuclear ${\rm Li}$ dimers will be subject to reactive 
trap losses, but all the remainder bi-alkali heteronuclear molecules 
should be stable with respect to atom exchange collisions 
in their ground rovibronic 
state~\cite{hutson-2010}. Of the remaining stable bi-alkali heteronuclear 
molecules $({\rm NaK, NaRb, NaCs, RbCs}),$ one of the best candidates 
for the observation of many-body effects caused 
by long-ranged dipolar interactions is ${\rm Na K}$, 
which can be fermionic ${\rm ^{23}Na ^{40}K}$ or 
bosonic ${\rm ^{23}Na ^{39}K}$ 
in nature. These systems are currently being pursued by some 
groups~\cite{NaK-molecules-1, NaK-molecules-2}. 
Another serious candidate is bosonic ${\rm RbCs}$, which is also being 
explored experimentally~\cite{RbCs-molecules}. The formation of trimers
of the remaining stable heteronuclear molecules was also theoretically
found to be highly endoenergic for ground robrational singlet 
states~\cite{hutson-2010}, and it is also very likely to be endoenergic
for the first few excited robrational singlet states. 
Nevertheless, additional experimental and theoretical studies 
of few body effects like dimer, trimer and tetramer formation 
and stability need to be performed.
Although losses are expected due to attractive interactions 
necessary for pairing and superfluidity of fermionic dipolar molecules, 
it is not yet known theoretically or experimentally how big losses will be.
However, preliminary theoretical~\cite{hutson-2010} and 
experimental~\cite{NaK-molecules-1} work seem to suggest that 
fermionic ${\rm ^{23}Na ^{40}K}$ molecules are arguably the best candidate 
for the many body effects that lead to the emergence of 
fermionic dipolar superfluidity with breaking of time reversal symmetry, 
as discussed in the main text.


\begin{thebibliography}{99}
\bibitem{footnote1}
Although Strontium Ruthenate has been conjectured to be
a triplet $p_x \pm i p_y$ superconductor which breaks time reversal symmetry
spontaneously, experimental evidence is not conclusive since no
spontaneously circulating currents have been observed.
\bibitem{footnote2}
The inclusion of next-nearest neighbor terms representing 
the screened dipolar interactions reveal its anisotropic nature 
even in the cases of $\phi = \pm \pi/4, \pm 3\pi/4$,
and the four-fold symmetric superfluid phases discussed here may acquire 
a subdominant two-fold component.
\end{thebibliography}
\end{document}